\documentclass[aps,prb,twocolumn,amsmath,amssymb]{revtex4}
\usepackage{epsfig}
\usepackage{graphicx}

\begin{document}

\title{Paraconductivity of K-doped SrFe$_2$As$_2$ superconductor}

\author{P. Marra,$^{1}$ A. Nigro,$^{2,3}$  Z. Li,$^{4}$  G. F. Chen,$^{4}$ N. L. Wang,$^{4}$ J. L. Luo,$^{4}$ and C. Noce$^{2,3}$}

\affiliation{$^1$Institute for Theoretical Solid State Physics, IFW--Dresden, D01171 Dresden, Germany}
\affiliation{$^{2}$ SPIN-CNR, I-84084 Fisciano (Salerno), Italy}
\affiliation{$^{3}$ Dipartimento di Fisica ``E. R. Caianiello'' Universit\`a di
Salerno, I-84084 Fisciano (Salerno), Italy}
\affiliation{$^{4}$ Beijing National Laboratory for Condensed Matter Physics, Institute of Physics, Chinese Academy of Sciences, Beijing 100190, China}

\date{\today}

\begin{abstract}
Paraconductivity of the optimally K-doped SrFe$_2$As$_2$ superconductor is investigated within existing fluctuation mechanisms. The in-plane excess conductivity has been measured in high quality single crystals, with a sharp superconducting transition at T$_c$=35.5~K and a transition width less than 0.3~K. The data have been also acquired in external magnetic field up to 14~T. We show that the fluctuation conductivity data in zero field and for temperatures close to T$_c$, can be explained within a three-dimensional Lawrence-Doniach theory, with a negligible Maki-Thompson contribution. In the presence of the magnetic field, it is shown that paraconductivity obeys the three-dimensional Ullah-Dorsey scaling law, above 2~T and for H$\parallel$c. The estimated upper critical field and the coherence length nicely agree with the available experimental data.
\end{abstract}


\maketitle

\section{Introduction}
Layered Fe-As systems have attracted enormous interest because they have been identified as novel high-T$_c$ materials.  Indeed,  the family REO$_{1-x}$F$_x$FeAs (1111), where RE denotes a rare earth ion, shows transition temperatures T$_c$ up to 26~K~\cite{kamihara08,chen08a, zhu08, sefat08} for RE=La and even more high values of T$_c$ $\sim$ 41-55~K for Ce~\cite{chen08b}, Sm~\cite{chen08c}, Nd and PrO~\cite{ren08a,ren08b}, whereas the family AFe$_2$As$_2$  (122), where A is an Sr or Ba ion, exhibits a T$_c$ up to 38~K in doped BaFe$_2$As$_2$~\cite{} and similar critical temperatures in doped SrFe$_2$As$_2$.~\cite{rotter08,chen08d,sasmal08}

These materials are layered superconductors located on the border of magnetic instability as Cu-based compounds. Therefore, they should have an unconventional mechanism of superconductivity, which may be connected with magnetic fluctuations and/or spin density wave anomaly.~\cite{stewart11}

Since single crystals of these superconductors are available, intrinsic relevant physical properties can be investigated and noteworthy topics can be addressed providing a major breakthrough in the study of the electronic and superconducting  properties of Fe-As layered superconductors. In this paper, we investigate the nature of thermodynamical fluctuations above the superconducting critical temperature evaluating the resistivity in single crystal samples of a K-doped 122  superconductor. The motivation of this analysis is the following: it is known that near the transition region, thermodynamic fluctuations may give rise to an anomalous increase in the superconducting properties even at temperatures above T$_c$.~\cite{skocpol75} This fluctuation effect is very important because it may give valuable information on the superconductivity once physical quantities such as the conductivity, the magnetization, and the thermoelectricity are measured. Furthermore, it is of theoretical relevance since it may provide a stringent test of scaling theories in the critical region.~\cite{larkin05} Remarkably, the study of the excess conductivity above the superconducting transition has proved to be a powerful tool to investigate the dimensionality of the superconductivity. We would like also to stress that the identification of three-dimensional (3D) rather than two-dimensional (2D) thermal fluctuations turns out to be a crucial point for potential applications. For instance, the 2D nature of fluctuations in cuprates imply the occurrence of pronounced dissipation in the mixed state which is detrimental for applications.~\cite{vidal98}

However, few results are available on fluctuation conductivity in iron-based 122 superconductors. Here, we give a contribution to this issue, presenting a detailed analysis of the paraconductivity of doped Sr$_{0.6}$K$_{0.4}$Fe$_2$As$_2$ superconductor. We notice that this compound may be considered in the optimally-doped regime, since Sr$_{1-x}$K$_{x}$Fe$_2$As$_2$ exhibits the highest critical temperature for $x\sim$0.4-0.5.~\cite{sasmal08}

Referring to 1111 family, fluctuation conductivity studies have been carried out in 1111 Sm F-doped compound and have been consistently interpreted within a 2D nature of superconductivity.~\cite{pallecchi09} A systematic study of the excess conductivity in a similar compound, namely SmFeAsO$_{0.85}$ has been performed and 2D-3D crossovers have been detected.~\cite{solovjov10} We notice that a fluctuation conductivity investigation under magnetic field has been carried out also on REFeAsO (RE=Nd, Pr, Sm) superconductors again supporting a 2D-3D crossover of the paraconductivity.~\cite{liu10} Finally, in both electron- and hole-doped LaOFeAs compounds a paraconductivity study has been realized and it is found that the experimental data can be explained within a 3D AL theory.~\cite{liu11} For completeness, we would like to point out that all these results have been obtained using polycrystalline samples.

Concerning the 122 family, the excess conductivity has been measured for the optimally-doped BaFe$_{1.8}$Co$_{0.2}$As$_2$ single crystals with T$_c$=24.6~K. The analysis has shown that the superconductivity follows a 3D scaling rather than a 2D scaling even though the sample has a layered structure.~\cite{kim10} Very recently, it has been reported that
the overall resistivity for optimal, overdoped and underdoped Ba$_{1-x}$K$_{0.4}$Fe$_2$As$_2$ samples can be also described by a uniform formula taking into account the 3D fluctuation contribution.~\cite{liu11b}  Nevertheless, we would like to notice that the experimental data used in this case are taken from the literature, and the analysis is performed in the absence of the magnetic field.
The results here discussed have been obtained for high quality single crystals, and in the presence of an external magnetic field up to 14.~T. Our study suggests that Sr$_{0.6}$K$_{0.4}$Fe$_2$As$_2$ superconductor behaves as a 3D system. We notice that this conclusion is further supported by other experiments, including high-resolution angle-resolved photoemission spectroscopy and high field transport measurements.~\cite{arpes}  Therefore, considering the results reported on the Co- and K-doped BaFe$_2$As$_2$ superconductors and those here discussed, we may conclude that the paraconductivity of AFe$_2$As$_2$ superconductors follows a 3D fluctuation behavior, regardless of the hole or electron doping as well as of the alcaline A ion.

Regarding the magnetization fluctuations,we point out that a 3D fluctuation magnetization of the lowest Landau level type has been observed in Ba$_{1-x}$K$_x$Fe$_2$As$_2$ single crystals in a temperature window above the magnetic field dependent critical temperature.~\cite{salem09} We notice that these results have been obtained for two K contents $x=$~0.28 and $x=$~0.25. A similar trend has been observed in optimally-doped BaFe$_{1.8}$Co$_{0.2}$As$_2$ single crystals: the scaling form of the fluctuation magnetization observed for this sample indicates that the fluctuation magnetization follows a 3D scaling form in the critical fluctuation region for suitable applied magnetic field, suggesting that this iron pnictide material belongs to the class of 3D superconductors.~\cite{choi09}

The paper is organized as follows: in next section we describe the sample preparation of Sr$_{0.6}$K$_{0.4}$Fe$_2$As$_2$ single crystals, and we comment on the experimental results related to electric transport measurements. Then, we describe in section II the theoretical background behind the  paraconductivity in superconductors considering some of the available theoretical approaches. Sect. III contains the application of the previously mentioned theoretical models to the experimental data and a discussion of the results. Finally, the last section is devoted to a summary of the obtained results.

\section{Sample preparation and characterization}

Superconducting single crystals Sr$_{0.6}$K$_{0.4}$Fe$_2$As$_2$ were prepared by high-temperature solution method using FeAs as flux, following the procedure described in Ref.~\onlinecite{yan08}. Elemental Sr, K, and FeAs, in the ratio of 0.5:1:4, were put into an alumina crucible and sealed in a Ta crucible under 1.6 atm of argon gas. Then, the Ta crucible was sealed in a quartz ampoule, heated at 1150$^{\circ}$~C for 5 h, and cooled slowly to 800$^{\circ}$~C over 50 h. Platelike crystals with sizes up to 10mm$\times$5mm$\times$0.5mm could be obtained after breaking the alumina crucible. Scanning electron microscopy, with energy dispersive x-ray, and induction-coupled plasma analysis confirmed that the elemental composition of the crystal is Sr$_{0.6\pm\delta}$K$_{0.4\mp\delta}$Fe$_2$As$_2$ with $\delta\leq$0.02.

%
\begin{figure}
\begin{center}
\includegraphics[width=0.35\textwidth, angle=270]{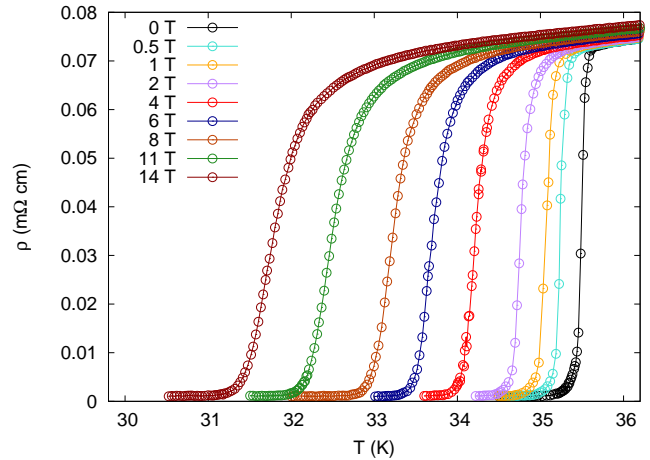}
\caption{Temperature dependence of the in-plane electrical resistivity, at fixed fields along $c$-axis, up to 14~T, from right (0~T) to left (14~T), in a temperature range near the superconducting transition.}
\end{center}
\end{figure}
%

The in-plane resistivity, measured by using a standard four-probe method, decreases with decrasing temperature and shows a downward curvature consistent with the polycrystal sample. With further decreasing temperature, an extremely sharp superconducting transition is observed at T$_c$=35.5~K, with a transition width less than 0.3~K, indicative of an high degree of homogeneity of the sample.~\cite{sasmal08,chen08d}

Figure 1 shows the temperature dependence near T$_c$ of dc resistivity in an external magnetic field. The magnetic field is applied along $c$-axis and its intensity varies between 0 and 14~T. It can noticed that, at the critical temperature, the resistivity shows a broadening, which is quite different from a conventional superconductor indicating the relevance of the fluctuations. The broadening is not large as in high temperature cuprate superconductors, but it is quite pronounced. Moreover, T$_c$ onset and the temperature of zero resistivity also decrease as the magnetic field is increased. Figure 3 shows the critical temperature T$_c$(H) as function of the applied magnetic field. The error bars have been determined from the 10$\%$ and 90$\%$ resistance drop at T$_c$(H) criterion.
For completeness, we notice that this behavior is different from polycrystalline LaFeAsO samples where the superconducting transition is strongly broadened in a magnetic field.~\cite{chen08a}

%
\begin{figure}
\begin{center}
\includegraphics[width=0.45\textwidth]{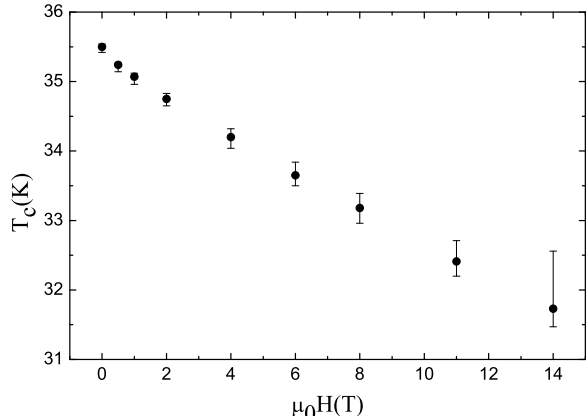}
\caption{Critical temperature as function of the applied magnetic field. Error bars correspond  to the transition width evaluated from the 10$\%$- 90$\%$ criterion.}
\end{center}
\end{figure}
%

\section{Theoretical framework}

The measured conductivity $\sigma (T)$ can be written as:
\[
\sigma (T)=\sigma_n(T)+\Delta \sigma\,
\]
where $\sigma_n(T)$ is the temperature dependent normal state dc conductivity and $\Delta \sigma$, the so-called paraconductivity, is the additional conductivity due to fluctuation effects near the superconducting transition.

\noindent Aslamazov and Larkin (AL)~\cite{aslamazov68} pointed out that thermal fluctuations in a superconductor result in a finite probability of a Cooper pair formation above T$_c$, leading to an excess electrical conductivity for T$>$T$_c$. The fluctuation enhanced conductivity is given in 3D and 2D by, respectively,
\begin{equation}
\Delta \sigma^{AL}_{3D}=\frac{e^2}{32\hbar}\frac{1}{\xi(0)}\varepsilon^{-1/2}
\end{equation}
and
\begin{equation}
\Delta \sigma^{AL}_{2D}=\frac{e^2}{16\hbar}\frac{1}{d}\varepsilon^{-1}
\end{equation}
where $\xi(0)$ is the coherence length at zero temperature, $\varepsilon=\ln(T/T_c)\approx(T-T_c)/T_c$, and $d$ is the characteristic length of the 2D system, i.e. the thickness in superconducting films or the distance between adjacent layers in layered superconductors.

In order to express the 2D-3D crossover regime, Lawrence and Doniach (LD)~\cite{lawrence71} modeled layered superconductors as stacks of 2D superconducting layers, weakly coupled by Josephson tunneling. The explicit expression of this quantity within the LD model is given by
\begin{equation}
\Delta \sigma^{LD}=\frac{e^2}{16\hbar d}(1+2 \alpha)^{-1/2} \varepsilon^{-1}
\end{equation}
where $d$ is the distance between adjacent layers and $\alpha$ is a coupling parameter given by:
\begin{equation}
\alpha=\frac{2\xi_c^2(T)}{d^2}=\frac{2\xi_c^2(0)}{d^2\varepsilon}
\end{equation}
Here, $\xi_c(0)$ is the zero-temperature $c-$axis coherence length.

We point out that in this case the expression of the paraconductivity models the crossover from the 2D limit at high temperature to the 3D behavior as the temperature-dependent coherence length $\xi_c(T)$ along the stacking direction exceeds the layer distance at temperature near T$_c$ . In other words, Eq.~(3) would be reduced to the 2D AL form Eq.~(2) for T $\gg$ T$_c$, corresponding to $\xi_c(T)\ll d$, and to the 3D AL form Eq.~(1) for T $\sim$ T$_c$, i. e. $\xi_c(T)\gg d$ and it exhibits a 2D-3D crossover at $T^{\star}\approx T_c \{1+[2 \xi_c(0)/d]^2\}$, that corresponds to the limit $\xi_c(T)\sim (d/2)$.

The above theoretical models are based on direct contribution to excess conductivity, i.e. contributions due to the acceleration of Cooper pairs generated by thermodynamic fluctuations. An indirect contribution
to excess conductivity arises from the interaction of fluctuating Cooper pairs with normal electrons
and was calculated by Maki~\cite{maki68} and later modified by Thompson.~\cite{thompson70}
The indirect anomalous contribution to excess conductivity for layered superconductors was derived by Hikami and Larkin~\cite{hikami88}, and independently by Maki and Thompson.~\cite{maki89} In the absence of applied magnetic field, it is given by
\begin{equation}
\Delta \sigma^{MT}=\frac{e^2}{8\hbar d}\frac{1}{(\varepsilon-\delta_{MT})} \ln \left ( \frac{\delta^{\ast}} {\alpha}
\frac{ 1+\alpha+\sqrt{1+2 \alpha} }{ 1+\delta^{\ast}+\sqrt{1+2 \delta^{\ast}} }
\right )
\end{equation}
where $\alpha$ is given by Eq.~(4), $\delta_{MT}= \pi \hbar/8 k_B T \tau_{\phi}$, $\delta^{\ast}=16\xi_c^2(0)k_B  T \tau_{\phi}/\pi \hbar d^2$, $\tau_{\phi}$ being the phase-relaxation time of the quasiparticle.

\noindent In the three-dimensional case, when $\delta_{MT}\ll \epsilon \ll \epsilon \alpha$, Eq.~(5) reduces to
\[
\Delta \sigma^{MT}_{3D}=\frac{e^2}{8\hbar \xi_c(0)}  \left ( \sqrt{\delta_{MT}} + \sqrt{{\epsilon}} \right )^{-1}
\]
\noindent while in the two-dimensional limit, where $\delta_{MT} \gg \epsilon \alpha$ and $1 \gg \alpha$, we have
\[
\Delta \sigma^{MT}_{2D}=\frac{e^2}{8\hbar d}   \frac{1}{(\varepsilon-\delta_{MT})} \ln \left (\frac{\epsilon }{\delta_{MT} } \right )
\]

\noindent We point out that Eq.~(5) is valid only for $\delta_{MT} \leq 0.1$.~\cite{thompson70,maki89,keller72}

The presence of an applied magnetic field raises the complexity of the analysis of fluctuation conductivity. Indeed, the quantity $\varepsilon=(T-T_c(H))/T_c(H)$ not only strongly depends on the temperature, but also on the magnitude of the magnetic field. In particular, in a magnetic field sufficiently strong, the paired quasiparticles are confined to the lowest Landau level (LLL) and the transport is limited along the field direction.
Within the LLL approximation, Ullah and Dorsey (UD) calculated the fluctuation conductivity including the free energy quartic term within the Hartree approximation, obtaining a scaling law in a magnetic field, in terms of unspecified scaling functions $f_{3D}$  and $f_{2D}$, valid for 3D and 2D superconductors, respectively.~\cite{ullah90} The results are
\begin{equation}
\Delta \sigma^{UD}_{3D} (H)=\left [ \frac{T^2}{H} \right ]^{1/3}
f_{3D} \left [B \frac{T-T_c(H)}{(TH)^{2/3}} \right ]
\end{equation}
for a 3D system, and
\begin{equation}
\Delta \sigma^{UD}_{2D} (H)=\left [ \frac{T}{H} \right ]^{1/2}
f_{2D} \left [ A \frac{T-T_c(H)}{(TH)^{1/2}} \right ]
\end{equation}
for a 2D system.

\noindent Here $A$ and $B$ are appropriate constants characterizing the materials. Such scaling behaviors have been found in high temperature superconductors above a characteristic field of the order of few Tesla,~\cite{welp91,menegotto97} and more recently, in the polycrystalline samples of SmFeO$_{1-x}$F$_x$ compound, with a measured field $\mu_0$ H$_{LLL} =$8~T,~\cite{pallecchi09} and in FeTe$_{1-x}$S$_{x}$ and in FeSe$_{0.9-x}$(Si,Sb)$_x$ superconductors, with a measured field $\mu_0$ H$_{LLL}$ =6-8~T.~\cite{pandya10} We notice that H$_{LLL}$ is the field above which only few Landau levels are occupied with negligible inter-Landau level interactions, so that the LLL approximation holds.

\section{Results and discussion}
In this section we firstly apply the AL, LD  and the MT approaches to investigate the dimensionality of the superconductivity of Sr$_{0.6}$K$_{0.4}$Fe$_2$As$_2$, then we study the effect of the magnetic field on the conductivity fluctuations by means of the Eqs.(6)-(7).

To evaluate the paraconductivity, we follow the procedure outlined in Ref.~\onlinecite{kim89}. In this paper it has been shown that a better fit may be obtained evaluating the conductivity $\sigma(T)$
\[
\sigma(T)=\frac{1}{\rho_n(T)}+ \Delta \sigma(T)
\]
over the entire range of the fit. Here $\rho_n(T)$ is the normal state resistivity.

Therefore, we assume that the fluctuations contribute to $\sigma(T)$ over the entire temperature range of the fit.
To best fit the experimental data, we assume that:
\begin{equation}
\rho_n(T)=a T^2+ b T + c\, ,
\end{equation}
and within the AL theory
\[
\Delta \sigma(T)= K \varepsilon^{-\alpha} \quad ,
\]
\noindent where $\varepsilon=(T-T_c)/T_c$, $K={e^2}/{32\hbar \xi(0)} ({e^2}/{16\hbar d})$ and $\alpha=0.5 (1)$ in three (two) dimensions. We assume as fitting parameters $a$, $b$ and $c$ and $d$ ($\xi(0)$)  in 2D (3D), and for T$_c$ the value 35.5 K, corresponding to the temperature where the derivative of the resistivity curve $\rho(T)$ is maximum.

The best fit of our data, not reported here, gives the following results: in 2D the parameters entering the normal state resistivity are $a=$ 3.9 10$^{-6}$ m$\Omega$ cm K$^{-2}$, $b=$ 1.2 10$^{-3}$ m$\Omega$ cm K$^{-1}$ and $c=$ 2.6 10$^{-2}$ m$\Omega$ cm, while $K$ is equal to 1.1$\times$10$^{-3}$, from which we infer the following value for the characteristic length of the 2D system $d\simeq$ 1300~{\AA}. Performing the same analysis for the 3D case we obtain $a=$ 4.0 10$^{-6}$ m$\Omega$ cm K$^{-2}$, $b=$ 1.2 10$^{-3}$ m$\Omega$ cm K$^{-1}$ and $c=$ 2.7 10$^{-2}$ m$\Omega$ cm and  $\xi(0)=$ 36~{\AA}.
Although the application of the 3D limit of the
AL theory to our data gives a fit analogous to the one obtained in the 2D case, the $\xi(0)$ value inferred
from this calculation is in good agreement with the available experimental data obtained for this quantity for other 122 compounds, whereas the estimation of $d$ gives a rather unfair value.
Hence, these results suggest that the experimental data follow a 3D-like scaling law, and the coherence length is of the same order of magnitude of Ba-122 superconductors.~\cite{welp09}

%
\begin{figure}[!b]
\begin{center}
\includegraphics[width=0.5\textwidth]{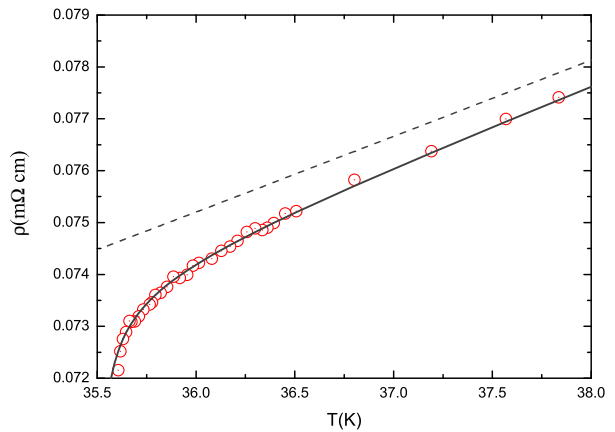}
\end{center}
\caption{ Resistivity plotted (circles) as a function of the temperature. The fit curve is represented by a full line and it has been obtained by means of the LD formula Eq.~3. We have also plotted the normal state resistivity with a dotted line.}
\end{figure}
%

Now, let us investigate the possibility of a 2D-3D crossover considering the LD approach for layered superconductors including the indirect MT term to the excess conductivity. We have fitted the experimental data using a more general formula where the MT contribution given by Eq.~5 is added to the LD formula Eq.~3. The best fit to our data is reported in Fig.~3, where we plot the experimental data together with the fit (full line) and the normal state conductivity (dotted line). From the fit we have obtained $d\approx$ 10~{\AA} and  $\xi_c(0)=$ 32~{\AA}. This means that our previous hypothesis about the 3D behavior of the scaling law is further supported by this value of the coherence length, that agrees with the one derived within the AL theory, and by the value of characteristic length $d$ which is of the same order of the interlayer distance. From the fitted value of $\delta^{\ast}$, the phase breaking time $\tau_{\phi}$  is estimated to be $\sim$ 10$^{-15}$ s at the onset of the superconducting transition.
This value of $\tau_{\phi}$ leads to an anomalous MT term lower than 10$\%$ of the total paraconductivity, in agreement with the irrelevance of this term. We note that the extremely low value of the pair breaking time here obtained is comparable to the values reported for high temperature superconductors.  In particular, it is incompatible with the condition $\delta_{MT} \leq 0.1$  under which the Eq.~(5) has been derived, requiring for the K-doped 122 compound a pair breaking time $\tau_{\phi} \geq 8 \cdot 10^{-13}$s. This is indicative that the indirect terms gives a negligible contribution to zero field excess conductivity, thus suggesting, as for high temperature superconductors, a non conventional mechanism for superconductivity in the Fe-based 122 materials.

Then, we analyze the fluctuation conductivity of the sample in the presence of the magnetic field by considering the 2D or 3D scaling behaviors described by Eqs.~(6)-(7). We notice that the critical temperatures used in the scaling of experimental data depend on the applied magnetic field as shown in Fig.~2.

%
%
\begin{figure}
\begin{center}
\includegraphics[width=0.45\textwidth]{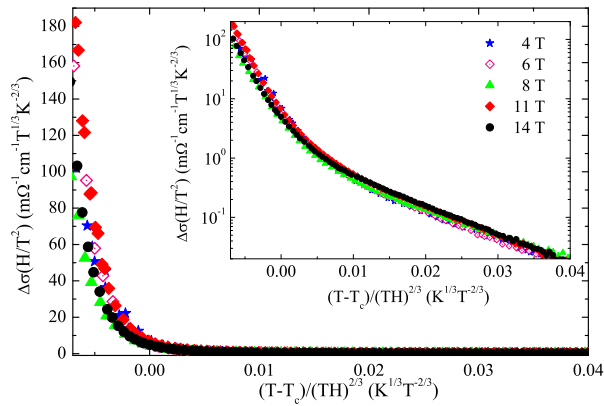}
\caption{Scaling plots of the excess conductivity plotted as a function of the variable $(T-T_c(H))/(TH)^{2/3}$ in magnetic fields $H$=4, 6, 8, 11 and 14 T, for the 3D model of the paraconductivity described by Eq.~6. The inset refers to the same data plotted in a semi-logarithmic scale.}
\end{center}
\end{figure}


In Fig.~4, the scaled excess conductivity $\Delta\sigma^{UD}_{3D}(H)$ for Sr$_{0.6}$K$_{0.4}$Fe$_2$As$_2$  single crystal is plotted for the case of 3D scaling; the same data are shown on a semi-logarithmic scale in the inset of this figure. For fields $\mu_0$H $\geq$ 4T, the data exhibit a very nice scaling behavior of the fluctuations above T$_c(H)$ up to the vanishing paraconductivity region ([T-T$_c(H)$/T$_c(H)$] $\leq$0.1), and also in an extended region below T$_c(H)$ down to temperatures where the resistivity  $\rho$ is approximately equal to 5$\% $ of the normal state resistivity $\rho_n$, at the onset of the superconducting transition. Below these temperatures, the dissipative effects of the vortex motion start to become important.~\cite{palstra90} We point out that the 3D scaling behavior has been obtained from the excess conductivity  $\Delta \sigma_H(T)$ curves by using T$_c(H)$ values deduced from the $d\rho/dT$ peak, {\it without using any other fitting parameters}. We would like to stress that a deviation from the scaling formula is observed for fields lower than 2~T. This seems to indicate that in our sample the fluctuation conductivity, at sufficiently high fields, is well described within the 3D lowest Landau level approximation and that the $\mu_0 H_{LLL}$ field is of the order of 2~T for  the H$\parallel$c configuration.

In Fig.~5, the 2D scaling is presented with the same T$_c(H)$ values and temperature range used for the 3D scaling. As it can be clearly seen from the inset of this figure, the 2D scaling plot gives rather poor results. We would like also to point out that we tried to apply a different fitting procedure, namely treating  T$_c(H)$ as a fitting parameter. The obtained results, not presented here, clearly show that also in this case the experimental curves do not merge into a single one.

The 3D scaling analysis gives a linear behavior of the upper critical field H$_{c2}^{\parallel c}$(T) at high fields with a slope  $\mu_0$dH$_{c2}^{\parallel c}$(T)/dT$\simeq$ 4.2~T/K. The in plane-upper critical field H$_{c2}^{\mid \mid ab}$(T) has been also measured on the same single crystal looking at the temperature dependence of the resistivity for applied fields up to 14~T and perpendicular to the $c$-axis. According to the Ginzburg-Landau formula $\xi_{ab,c}=\left [ \phi_0 T_c/2\pi \mu_0  \quad d\left (H_{c2}^{\parallel c,ab}
\right )/{dT}  \right ]^{1/2}$, the H$_{c2}^{\parallel c}$(T) and H$_{c2}^{\parallel ab}$(T) slopes at T$_c$ gives for the in-plane and out-of-plane coherence length of the sample the values $\xi_{ab}(0)$ = 21{\AA} and   $\xi_{c}(0)$ = 16{\AA} respectively, which agree well with different estimations of the same quantities for the Ba-122 compound.~\cite{welp09} Furthermore, the $\xi_{c}(0)$ value here deduced within the paraconductivity analysis in zero field is consistent, within a factor 2, with the value previously estimated.

As final remark, we point out that the presence of a sufficiently strong applied magnetic field reduces the effective dimensionality of the fluctuations, making them significant in a temperature window of the order of Gi(H)T$_c(H)$. Here Gi(H) is the field-dependent Ginzburg number: Gi(H)=(H/H$_{c2}$(0))$^{2/3}$ Gi(0)$^{2/3}$, H$_{c2}$(0) being the zero temperature c-axis upper critical field. We recall that Gi(0) defines the critical region $\mid$T-T$_c$$\mid$ $\ll$ Gi(0) $\times$ T$_c$ near the critical temperature where the fluctuations are significant. It is known that Gi(0) is $\sim$ 10$^{-9}$ for conventional superconductors and $\sim$ 10$^{-2}$ for high-temperature superconductors, while it is $\sim$ 10$^{-2}$ for Nd-1111 compound, and $\sim$ 10$^{-4}$  for Ba-122 superconductors.~\cite{welp09,putti10}
Our experimental results indicate that the actual temperature region of fluctuations is slightly larger
than the above-reported estimate, and also than the value obtained for the Ba-122 compound.

%
\begin{figure}
\begin{center}
\includegraphics[width=0.45\textwidth]{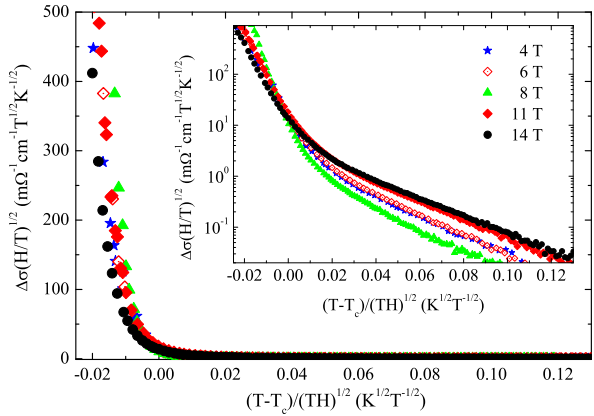}
\caption{Scaling plots of the excess conductivity plotted  as a function of the variable $(T-T_c(H))/(TH)^{1/2}$ in magnetic fields $H$=4, 6, 8, 11 and 14 T, for the 2D model of the paraconductivity described by Eq.~7. The inset refers to the same data plotted in a semi-logarithmic scale.}
\end{center}
\end{figure}

\section{Conclusions}

To summarize, we have measured the resistivity on high quality optimally K-doped SrFe$_2$As$_2$ single crystal superconductor in the absence and in the presence of an external magnetic field up to 14~T. We have presented a study of fluctuation conductivity in the Gaussian regime close and above T$_c$, also considering the presence of the external magnetic field. The data have been analyzed in terms of the theories for layered superconductors, including the AL, LD and MT models, and the UD approach when an external magnetic field is applied to the superconductor. At zero magnetic field, the analysis of the excess fluctuation conductivity shows that the superconductivity follows a 3D fluctuations behavior rather than a 2D one even though the superconductor has a layered structure.
Furthermore, the inclusion of the MT term in the analysis of the paraconductivity data, leads to an extremely small value of the phase pair-breaking lifetime $\tau_{\phi}$, suggesting a possible non conventional superconductivity in the Fe-based materials, as it happens in high temperature superconductors.

In presence of the magnetic field, the conductivity curves obey the 3D lowest Landau level scaling for fields above  $\approx$ 2T in the H$\parallel$c configuration. The measured $\rho(T)$ in different applied magnetic fields allows to extract the upper critical fields H$^{\parallel c}_{c2}$(T) and H$^{\parallel ab}_{c2}$(T), whose high field slopes are as large as -4.2~T/K and -8.1~T/K, respectively.  Moreover, the zero temperature out-of-plane coherence length deduced by the excess conductivity analysis, in zero field, turns out to be  $\xi_c(0)$ = 32~{\AA} which is of the same order of magnitude of the value obtained by the upper critical field data.

\end{document}